\documentclass[journal]{IEEEtran}
\usepackage[columnwise]{lineno}
\usepackage{cite}
\usepackage{balance}
\ifCLASSINFOpdf
  \usepackage[pdftex]{graphicx}
\else
  \usepackage[dvips]{graphicx}
\fi
\usepackage{graphicx}
\usepackage{amsmath}
\usepackage{amssymb}
\usepackage{mathrsfs}
\graphicspath{{images/}}
\usepackage{makecell}
\usepackage{subfig}
\usepackage{bm}
\usepackage{url}
\usepackage{hyperref}
\hypersetup{
 colorlinks=True,
}
\usepackage{booktabs}
\usepackage{makecell}
\usepackage{multirow}

\newcommand{\cmmnt}[1]{}

\usepackage[table,xcdraw, dvipsnames]{xcolor}
\usepackage[table,xcdraw]{xcolor}

\begin{document}
\title{SpectNet : End-to-End Audio Signal Classification using Learnable Spectrogram Features}

\author{Md.~Istiaq~Ansari, 
        Taufiq~Hasan,~\IEEEmembership{Senior Member,~IEEE}
        \thanks{Md. I. Ansari and T. Hasan are with the mHealth Laboratory, Department of Biomedical Engineering, Bangladesh University of Engineering and Technology (BUET), Dhaka 1205, Bangladesh (e-mail: \url{ansariistiaq@gmail.com}; \url{taufiq@bme.buet.ac.bd}).}
}
\markboth{IEEE ~Vol.~xx, No.~xx, xxxx~2021}
{Ansari \MakeLowercase{\textit{et al.}}: SpectNet : End-to-End Audio Signal Classification Using Learnable Spectrograms}

\maketitle

\begin{abstract}
Pattern recognition from audio signals is an active research topic encompassing audio tagging, acoustic scene classification, music classification, and other areas. Spectrogram and mel-frequency cepstral coefficients (MFCC) are among the most commonly used features for audio signal analysis and classification. Recently, deep convolutional neural networks (CNN) have been successfully used for audio classification problems using spectrogram-based 2D features. In this paper, we present SpectNet, an integrated front-end layer that extracts spectrogram features within a CNN architecture that can be used for audio pattern recognition tasks. The front-end layer utilizes learnable gammatone filters that are initialized using mel-scale filters. The proposed layer outputs a 2D spectrogram image which can be fed into a 2D CNN for classification. The parameters of the entire network, including the front-end filterbank, can be updated via back-propagation. This training scheme allows for fine-tuning the spectrogram-image features according to the target audio dataset. The proposed method is evaluated in two different audio signal classification tasks: heart sound anomaly detection and acoustic scene classification. The proposed method shows a significant 1.02\% improvement in MACC for the heart sound classification task and 2.11\% improvement in accuracy for the acoustic scene classification task compared to the classical spectrogram image features. The source code of our experiments can be found at \url{https://github.com/mHealthBuet/SpectNet}
\end{abstract}



\IEEEpeerreviewmaketitle

\section{Introduction}
\IEEEPARstart{A}{udio} pattern recognition is a prime research topic in machine intelligence. While computer vision aids intelligent systems to understand the visual world, audio pattern recognition automates the analysis of the world of sound. Audio pattern recognition encompasses various sub-areas including acoustic scene classification \cite{barchiesi2015acoustic,gharib2018acoustic}, music classification \cite{mckinney2003features, choi2017transfer}, speech emotion classification \cite{xie2019speech}, and sound event detection \cite{portelo2009non}. Audio analysis can also be applied in the healthcare domain, for example in pathological speech detection \cite{panek2015acoustic}, heart \cite{heart_sound_old_work,zabihi,potes,prio_vai} and lung \cite{lung_sound1,lung_sound2} sound analysis.

\begin{figure}
    \centering 
    \includegraphics[width=\linewidth] {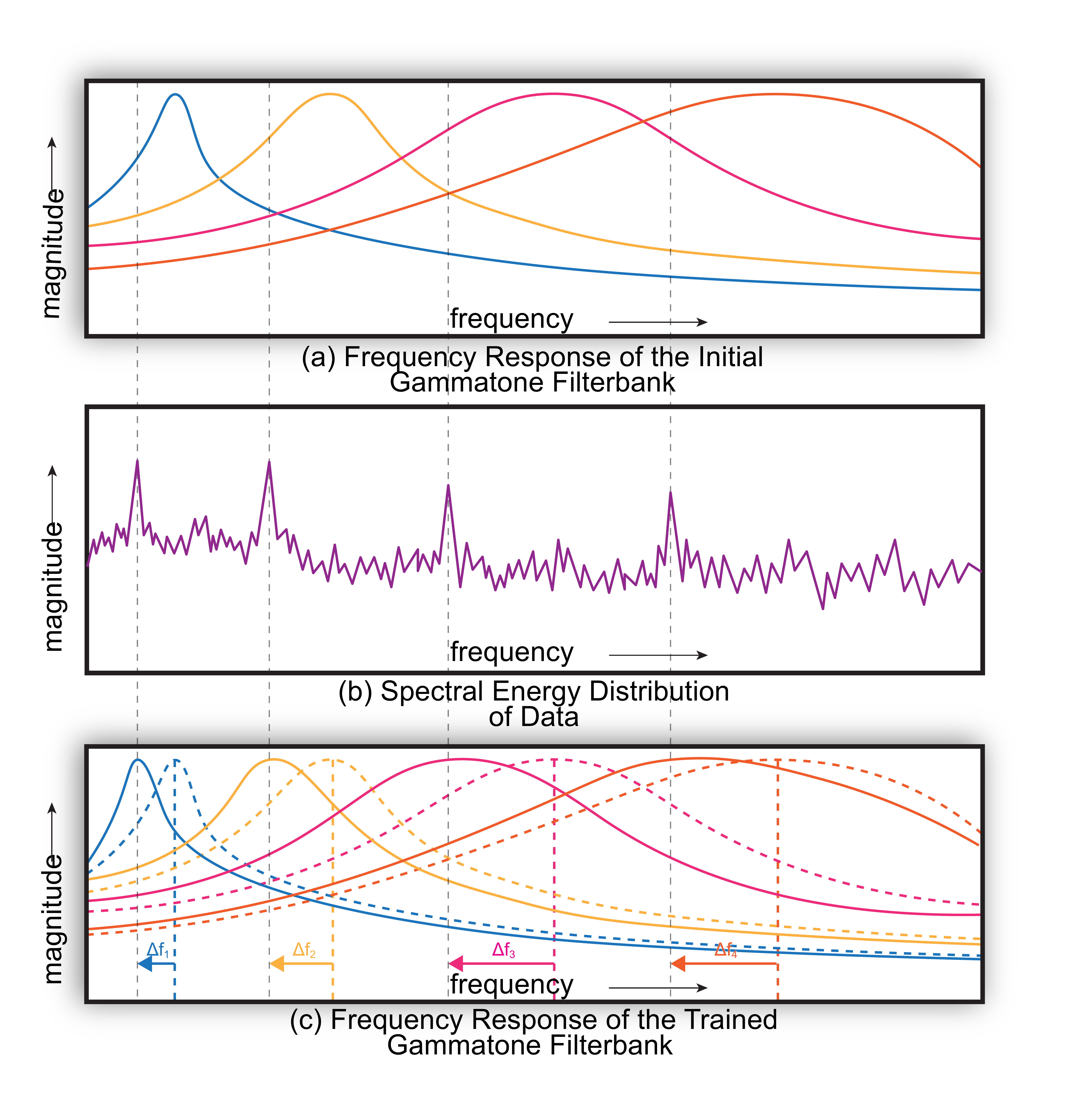}
    \caption{A conceptual illustration of a data-driven learnable filterbank before and after training. (a) Initial frequency response of a gammatone filterbank, (b) Spectral energy distribution of observed data, (c) Frequency response of gammatone filterbank after learning from the data.}
    \label{fig:learning_effect}
\end{figure}

A considerable amount of research has been done on various domains of audio signal analysis. Early works were limited to private datasets, such as \cite{audio_old_work1} where a Hidden Markov Model (HMM) model was used to classify three types of sounds. In recent years, the audio signal classification topic has received increased attention with the availability of public datasets provided by the Detection and Classification of Acoustic Scenes and Events (DCASE) \cite{dcase2016_dataset_paper} challenges. Audio scene classification and audio tagging studies have been done using wavelet \cite{pann_large_audio} and spectrogram \cite{dcase_submission_70} features with deep neural networks. For an overview of the audio/acoustic scene analysis topic, the reader may refer to the following review articles \cite{portelo2009non,gharib2018acoustic}. In the healthcare domain, heart sound or the phonocardiogram (PCG) is one of the most commonly used biomedical signals for the early diagnosis of heart diseases. The public release of the Physionet heart sound dataset \cite{physionet_dataset_paper} has further stimulated research studies on this topic. A considerable amount of work has been done on PCG analysis using 1D-CNN \cite{potes}, learnable filterbank \cite{prio_vai}, and time-frequency features \cite{zabihi}. Heart sound classification using spectrogram feature has been explored with SVM models \cite{heartsound_spectrogram_svm}, partial least squares regression models \cite{heartsound_spectrogram_reggress}, Laconic Neural Network \cite{heartsound_spectrogram_LNN}, LSTM network \cite{heartsound_spectrogram_lstm} and Convolutional Neural Networks \cite{heartsound_spectrogram_cnn,heartsound_spectrogram_cnn2}. MFCC features have been experimented with dynamic time wrapping for classification \cite{heartsound_dtw_mfcc} and CNN-LSTM models \cite{heartsound_mfcc_cnnlstm}

A brief inspection of the previous research on audio analysis quickly reveals the importance of spectrogram or short-term spectrum analysis-based acoustic features. Although first proposed for speech signal processing, the use of MFCC \cite{davis1980comparison} features seem to become virtually ubiquitous for any audio signal \cite{gharib2018acoustic,mckinney2003features, xie2019speech, portelo2009non, panek2015acoustic}. Short-time Fourier transform (STFT) is one of the most frequently used pre-processing techniques used for audio/speech signal analysis\cite{stft}. STFT is fundamental to extracting many other spectral features, including MFCCs and even segment-level features such as the i-vector, commonly used in speaker and language recognition \cite{ivector_taufiq,ivector2}. With the advancement of deep learning algorithms for image classification, convolutional neural networks (CNN), particularly the 2D representation of 1-D signals, have become popular among researchers. Spectrogram and MFCC features are the most common time-frequency features used for such tasks. These features along with CNN classifiers have performed well in various domains of audio and 1-D signal classification including, acoustic scene classification \cite{dcase_super_vector}, ECG classification \cite{huang2019ecg}, speech emotion recognition \cite{zhao2019speech} and other biomedical signal classification tasks \cite{heartsound_dtw_mfcc,heartsound_mfcc_cnnlstm}, emotion detection \cite{mfcc_emotion}.

The human auditory system is a highly effective transducer capable of distinguishing the subtle nuances of the audio signal. The motivation behind designing the mel-scaled filterbank structure used for MFCCs was based on the frequency response of the human auditory system \cite{cochlea}. In conventional mel-filter-bank analysis using gammatone filters \cite{holdsworth1988implementing}, the center frequencies for each filter is set to a fixed value calculated using the mel-scale (detailed in Sec. \ref{subsection_gammatone_filterbank}). 
The audio signal is passed through this fixed filterbank to generate a power-spectrum used for subsequent MFCC feature extraction or spectrogram analysis. However, different datasets may have a slightly different distribution of audio frequency peaks across the frequency axis, and filters with specific center-frequencies may attain an improved spectrogram representation and consequently extract superior features. Thus, a learnable spectrogram feature extraction scheme embedded in a deep learning architecture can improve classification performance on audio-based classification tasks.

In this work, we propose a front-end layer that generates 2D spectrogram image representation from raw audio waveforms using a set of learnable gammatone filters. The front-end layer can be integrated within a 2D-CNN model for end-to-end implementation. The proposed architecture provides a unique advantage of adjusting the filterbank center frequencies according to the specific classification task. The contributions of this work are as follows:
\begin{itemize}
    \item A front-end layer within a CNN architecture is proposed that provides a learnable spectrogram image representation from raw audio waveforms. The proposed layer can be integrated with any traditional 2D CNN model.
    \item Implementation details and mathematical formulations are provided for the proposed front-end layer in the case of a gammatone filterbank. 
    \item Experiments are performed on two different audio classification tasks, namely, acoustic scene classification and heart sound abnormality detection. The results show that the proposed spectrogram representation learning provides superior performance for both of the tasks when compared to fixed-parameter filterbank-based spectrogram image features.
\end{itemize}

The remainder of this paper is organized as follows. We explain our motivation of the SpecNet architecture in Sec. \ref{motivation}. This is immediately followed by Sec. \ref{proposed_method} that contains the background and the implementation details of the proposed filterbank layer along with the model architectures used for classification. Sec \ref{experiment} details the datasets, training regime, and evaluation setup used in the experimental evaluation and also elaborates the effect of our proposed layer on classification performance. Finally, we conclude and summarize our work in Sec. \ref{conclussion}.

\begin{figure}
\centering 
\includegraphics[width=\linewidth] {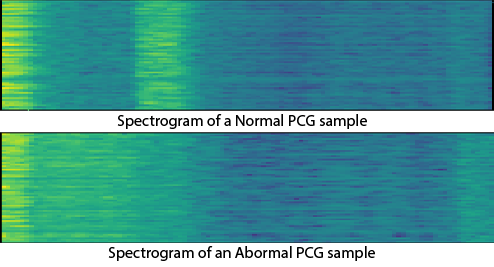}
\caption{Sample of Spectrogram image feature extracted from a segmented beat of PCG audio signal.}
\label{fig:spectrogram_features}
\end{figure}

\section{Motivation} \label{motivation}
Audio classification task has been explored using various machine learning methods in the past. CNNs have been the most popular method in this area as one of the best performing architectures \cite{cnn_svm}. In general, CNN models learn feature representations across the data dimension that is invariant to its location. While the 2D CNN may learn spatial invariant features on an image, the 1D CNN learns such features over the time axis for time-series data. Several state-of-the-art model architectures already demonstrate outstanding performance on various image classification tasks. Therefore, many researchers have recently begun classifying audio signals using state-of-the-art 2D CNN models by first converting the signals into 2D images, mainly using the spectrogram representation. 

The spectrogram is a 2D time-frequency feature representation that shows the audio signal's frequency components on Y-axis and temporal information on the X-axis. This 2D representation is commonly used as an input to a 2D CNN. The system can extract information from higher and lower frequency bands separately, searching for the pattern on the spatial representation of an audio signal. One of the most commonly used 2D time-frequency feature representations is generated using MFCCs\cite{mfcc} that was originally designed for speech. 

Inspired from this \cite{mfcc}, we consider the fact that different source domains of audio signals contain important information in different frequency bands. Thus, instead of manually setting the spectrogram frequency bands, learning to identify the most important frequency regions should improve classification performance. Following this assumption, we design a front-end filterbank layer that extracts spectrogram features while making the center frequency, bandwidth, and other filterbank parameters learnable from the data. In this way, the system can tune itself for the particular source domain of the signal. A conceptual demonstration of the idea is shown in Fig.\ref{fig:learning_effect}, where Fig.\ref{fig:learning_effect}(a) shows the initial frequency response of the filterbank consisting of four filters. Fig.\ref{fig:learning_effect}(b) shows the power spectrum of a hypothetical data frame that is to be classified. For ease of demonstration, we show here that the main lobe of the filters does not properly cover the peeks of the spectral distribution of the data. In other words, the filter attenuates the peak frequency regions where we presume the important information is most likely embedded. The proposed method is motivated by the idea that a learnable filterbank can shift the center frequencies towards the peaks (or an appropriate frequency region) by various amounts, as shown in Fig.\ref{fig:learning_effect}(c) so that the filters are more effective in extracting the relevant spectrogram features for subsequent classification. 

\section {{Proposed Method}} \label{proposed_method}
In this section, we describe the mathematical formulation and the implementation details of the proposed front-end layer for the 2D spectrogram representation learning framework and its integration with a 2D CNN architecture to classify audio signals.

\subsection{Background and Overview}

\subsubsection{Spectrogram}
The spectrogram visually represents a signal showing the energy variation on different frequency bands with time. Thus, we also refer to it as a time-frequency representation of a signal. The spectrogram of an audio signal can be generated using two main approaches. The first method involves using bandpass filters to divide the signal into different frequency bands. The second method uses the short-term Fourier transform (STFT). The spectrogram is generated from the signal by first segmenting it into chunks of very short-time windows. Next, the power spectrum is calculated using either of the methods mentioned. Each window results in a vertical plot that represents spectral energy distribution for that specific time window.

\subsubsection{Mel frequency cepstral coefficents (MFCC)}
The MFCC is the most popular short-term acoustic feature used for audio processing. The development of the MFCC features was inspired by the human auditory system's response to a different range of frequencies. Extraction of MFCC features includes segmentation of the audio into short overlapping frames followed by a windowing operation (e.g., Hamming). Next, the power spectrum is calculated in each short-time segment using Discrete Fourier Transform (DFT). The output is passed through a triangular-shaped filterbank. The frequency bins of the filterbank are calculated using the mel-scale \cite{davis1980comparison}. Logarithmic compression is applied to the output of each filter. Finally, Discrete Cosine Transform ($DCT$) is applied in order to decorrelate the features and perform compression.
\subsubsection{Gammatone filterbank} \label{subsection_gammatone_filterbank}
The proposed front-end layer in our system uses a gammatone filterbank-based implementation for spectrogram image extraction. The gammatone filter coefficients can be obtained as
\begin{equation}
    g(t) = at^{n-1} e^{-2\pi bt} \cos(2\pi f t+\sigma )
    \label{eqn:gammatone}
\end{equation}
where $a$ is the amplitude, $f$ is the filter's center frequency, $b$ denotes the bandwidth of the filter, $n$ is the filter order, and $\sigma$ is the phase of the gammatone wavelet (in radians). The center frequencies for each filter are calculated as follows. First, the minimum and maximum frequencies that define the frequency range of the data are converted into the mel-frequency scale using the equation below. 
\begin{equation}
    f_{mel} = 2595 \times { \log \left(1+ \frac{f_{hz}}{700}\right)}\label{eqn:hz2mel}
\end{equation}
The range is then divided into uniformly spaced $N$ values of frequencies where $N$ is the number of filters to be used. Finally, the frequency values from mel-scale are converted back to Hz unit using the following function. 
\begin{equation}
    f_{hz} = 700 \times 10^{\left(\frac{f_{mel}}{2595}-1\right)}\label{eqn:mel2hz}.
\end{equation}
To obtain a bandwidth equivalent to the rectangular filterbank bandwidth, the value of $b$ is set as
\begin{equation}
    \label{eqn:bandwidth_formula}
    b_k = 24.7\times\left(4.37\times\left(\frac{f_k}{1000}\right)+1\right).
\end{equation}
Overlapping windows are generally used for gammatone filterbank analysis. In our implementation, the gammatone filterbank coefficients are generated using a particular kernel size representing the time window where the signal is chunked to apply the filter.

\begin{figure}
\centering 
\includegraphics[width=\linewidth] {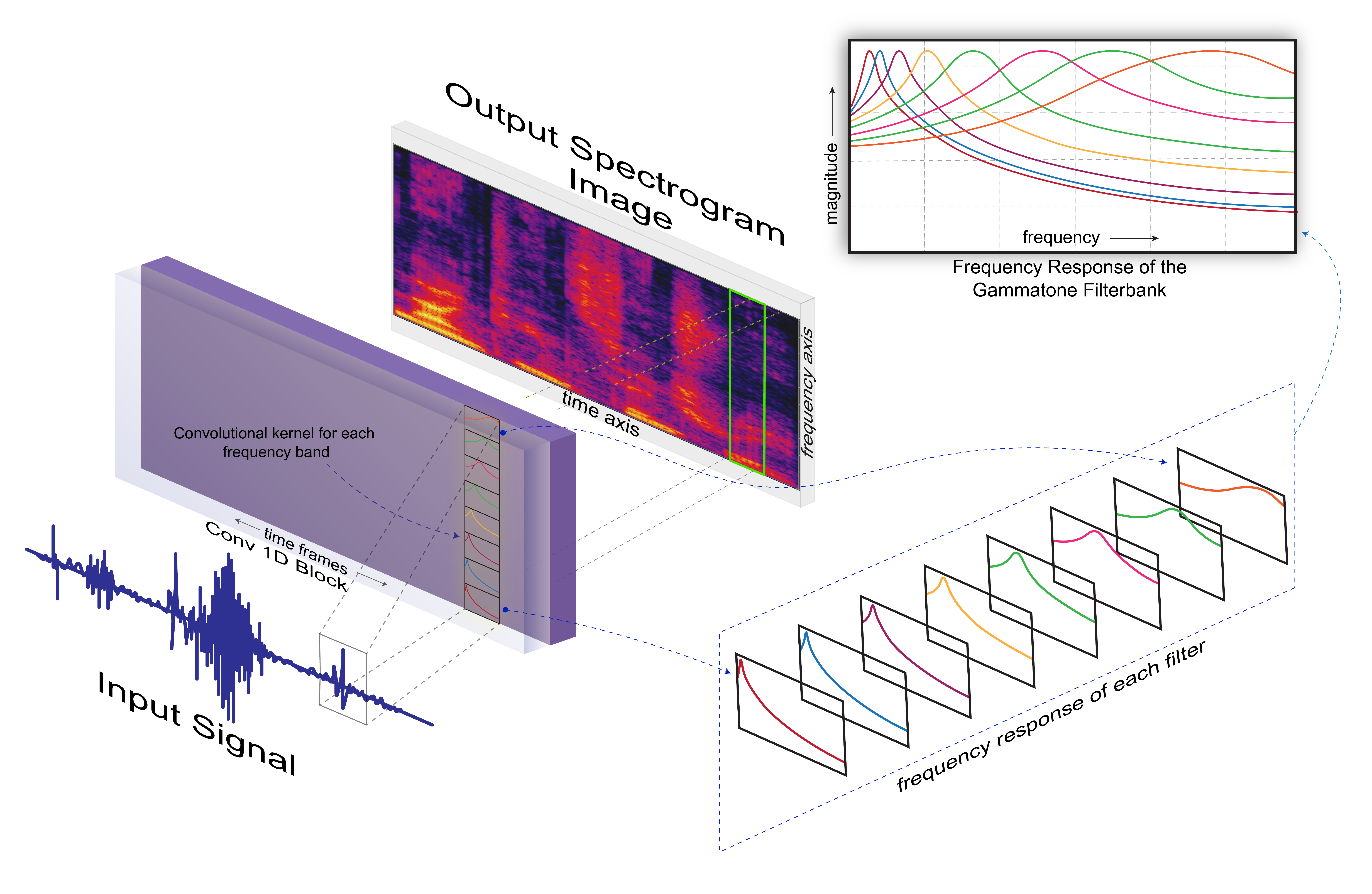}
\caption{Proposed front-end layer architecture for learnable spectrogram image feature extraction. The output of the layer can be fed into a traditional 2D CNN architecture for subsequent processing.}
\label{fig:frontendlayer}
\vspace{2mm}
\end{figure}

\begin{figure}
\centering 
\includegraphics[width=\linewidth] {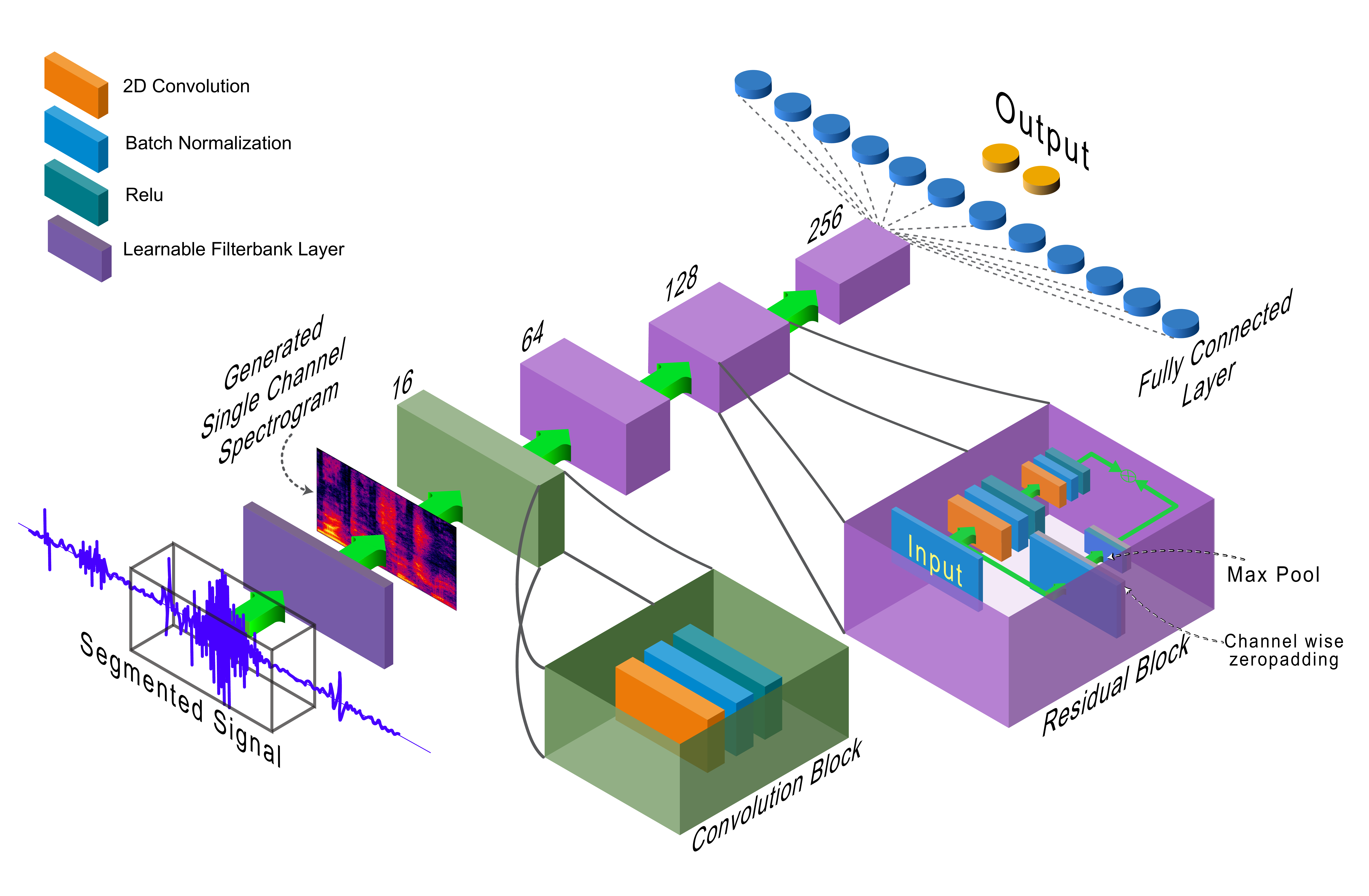}
\caption{The proposed SpectNet model architecture used for heart sound abnormality detection. The learnable spectrogram extraction layer is connected to a conventional 2D CNN architecture for classification.}
\label{fig:model}
\vspace{2mm}
\end{figure}

\subsection{Implementation of the learnable filterbank representation}
The proposed filterbank layer consists of some hyperparameters that need to be set before training. These include the number of filters used, the frequency range for analysis, window size, and overlap size. The frequency range should be set based on prior knowledge regarding the time-frequency characteristics of the relevant audio signal. The window size and overlap size should also be set depending on the temporal characteristics of the audio signal.

The proposed filterbank coefficients are generated using \eqref{eqn:gammatone}. The center frequencies of the filter bank are initialized with values used in a traditional mel-frequency filter bank. Therefore, initially, the layer extracts conventional mel-scale spectrogram image features. We hypothesize that using loss optimization through stochastic gradient descent, the learnable filter bank can extract center frequency parameters most relevant to the signal of interest. The learnable parameters of the filterbank are amplitude, filter order, bandwidth, and center frequency.

Let $\mathcal{L}$ be the loss function, $\mathbf{x} \epsilon \mathbb{R}^{1\times N}$ be a single channel input to the front-end layer, $\mathbf{g}_k \epsilon \mathbb{R}^{1\times K}$ be the $k$-th kernel, and $K$ denotes the kernel length and $\mathbf{z}_k$ its activation. The gradients for the parameters of the $k$-th filter are given by:
\begin{equation}
    \label{eqn:gammatonLoss}
    \frac{\partial \mathcal{L}} {\partial p_k} = \sum_{i=0}^{K-1}  \frac{\partial \mathcal{L}}{\partial g_k(i)} \frac{\partial g_k(i)}{\partial p_k}  \; \;    \text{for} \; \; p\, \epsilon \, \{ a,n,b,f \}
\end{equation}

Using \eqref{eqn:gammatone} and \eqref{eqn:gammatonLoss} the gradient for the center frequency, amplitude, bandwidth and order of the filter can be calculated respectively as:
\begin{align}
        \label{eqn:FcLossGammaton}
        \frac{\partial \mathcal{L}} {\partial f_k} &= \sum_{i=0}^{K-1}  \left(\frac{\partial \mathcal{L}}{\partial g_k(i)}\right) a_k t^{n_k-1} e^{-2\pi b_k t}[ - \sin(2\pi f_k t+\sigma ) 2\pi t ]
        \\
        \label{eqn:AmpLossGammaton}
        \frac{\partial \mathcal{L}} {\partial a_k} &= \sum_{i=0}^{K-1}  \left(\frac{\partial \mathcal{L}}{\partial g_k(i)}\right) t^{n_k-1} e^{-2\pi b_k t} \cos(2\pi f_k t+\sigma ) \\
        \label{eqn:BLossGammaton}
        \frac{\partial \mathcal{L}} {\partial b_k} &= \sum_{i=0}^{K-1}  \left(\frac{\partial \mathcal{L}}{\partial g_k(i)}\right) a_k t^{n-1} e^{-2\pi b_k t}[- \cos(2\pi f_k t+\sigma )2\pi t] \\
        \label{eqn:orderLossGammaton}
        \frac{\partial \mathcal{L}} {\partial n_k} &= \sum_{i=0}^{K-1}  \left(\frac{\partial \mathcal{L}}{\partial g_k(i)}\right) a_k t^{n-1} \ln{(t)} e^{-2\pi b_k t} \cos(2\pi f_k t+\sigma ) 
\end{align}
Given a single channel input $\mathbf{x} \epsilon \mathbb{R}^{1\times N}$ of length $N$, kernel/window size $K$ and a stride of $s$ since there is overlapping between window, the output length $M$ of a single filter can be calculated using:
\begin{equation}
    M = \left(\frac{N - K}{s}\right) + 1.
\end{equation}
Let the number of filters in the filterbank be $N_f$. Stacking up the output of each filter of length $M_x$ yields an output of shape $(M_x\times N_f$). This 2 dimensional representation of the gammatone filterbank output is fed to a 2 dimensinal CNN for feature extraction and classification. The front-end layer operation is shown on Fig. \ref{fig:frontendlayer}.

\renewcommand{\arraystretch}{1.5}
\begin{table}[t]
    \centering
    \caption{Data distribution of PhysioNet/CinC Challenge Database.}
    \resizebox{\linewidth}{!}{
        \begin{tabular}{@{}ccccc@{}}
            \toprule 
            \textbf{Subset} & 
            \textbf{\makecell{Total \\ Subject}} & 
            \textbf{\makecell{Normal \\ recordings}} & 
            \textbf{\makecell{Abnormal \\ recordings}} & 
            \textbf{Used device} \\\midrule
            a & 121 & 117 & 292 & Welch Allyn Meditron \\ 
            b & 106 & 385 & 104 & 3M Littmann E4000 \\
            c & 31 & 7 & 24 & AUDIOSCOPE \\
            d & 38 & 27 & 28 & Infral Corp. Prototype \\
            e (Norm.)  & 174 & 1867 & 0 & MLT201/Piezo \\
            e (Abn.) & 335\footnotemark & 0 & 151 & 3M Littmann \\
            f & 112 & 80 & 34 & JABES \\\bottomrule
        \end{tabular}}
    \label{data_physionet}
\end{table}
\footnotetext{Only 151 abnormal recordings from 'e' subset were used and due to data errors the rest were removed\cite{prio_vai}.}

\subsection{Model architecture}
The extracted spectrogram feature is represented as a 2D image and fed to a two-dimensional CNN model for classification. Since the learnable filterbank layer parameters depend on the audio signal characteristics and frequency range, we select the parameters according to the target domain. In this work, we performed experimental evaluations on two different tasks, namely, heart sound and acoustic scene classification. Accordingly, the model architecture had some notable differences for these two tasks. These task-specific design issues are discussed in Sec. \ref{architecture_design}. The model architecture used for heart sound classification is depicted in Fig. \ref{fig:model}.


\begin{table*}[!t]
    \centering
    \caption{Performance Comparison on Physionet Heart Sound Classification Task}
    \begin{tabular}{
>{\columncolor[HTML]{FFFFFF}}c 
>{\columncolor[HTML]{FFFFFF}}c 
>{\columncolor[HTML]{FFFFFF}}c 
>{\columncolor[HTML]{FFFFFF}}c 
>{\columncolor[HTML]{FFFFFF}}c 
>{\columncolor[HTML]{FFFFFF}}c 
>{\columncolor[HTML]{FFFFFF}}c }
\Xhline{2\arrayrulewidth}
Method               & Accuracy & F1    & Macc           & Sensitivity & Specificity & Precision \\ \Xhline{1\arrayrulewidth}
\multicolumn{7}{c}{\cellcolor[HTML]{EFEFEF}Baseline Methods}                                     \\
Gammatone 1D-CNN\cite{prio_vai}\footnotemark     & 75.80    & 83.17 & 82.30          & 91.30       & 73.29       & 76.36     \\
Gammatone 2D-CNN     & 76.69    & 84.67 & 84.03          & 92.03       & 76.03       & 78.39     \\
\multicolumn{7}{c}{\cellcolor[HTML]{EFEFEF}SpectNet (Static SpectNet)}                        \\
SpectNet-4 + ResNet  & 77.42    & 81.91 & 80.30          & 93.48       & 67.12       & 72.88     \\
SpectNet-8 + ResNet  & 81.64    & 87.27 & 87.66          & 86.96       & 88.36       & 87.59     \\
SpectNet-16 + ResNet & 81.79    & 87.37 & 87.68          & 87.68       & 87.67       & 87.05     \\
\multicolumn{7}{c}{\cellcolor[HTML]{EFEFEF}Proposed System (Learnable SpectNet)}              \\
SpectNet-16 + ResNet & 80.36    & 88.32 & \bf{88.70} & 87.68       & 89.73       & 88.97     \\ \Xhline{2\arrayrulewidth}
\end{tabular}
    \label{heartSoundTable}
\end{table*}
\footnotetext{The Heartnet model implemented in this experiment is taken from \cite{prio_vai}. To make the experiments with 1D and 2D features comparable, we used a 2D CNN with same depth as Heartnet. The 2D ResNet model is shown in Fig.\ref{fig:model}}

\section{Experiments and results} \label{experiment}
\subsection{Datasets}
To demonstrate the effectiveness of the proposed method, we evaluate it on two different audio datasets. These datasets include the 2016 PhysioNet/CinC challenge heart sound data \cite{physionet_dataset_paper} and the 2016 DCASE acoustic scene classification data \cite{dcase2016_dataset_paper}. The datasets are briefly described in the following sub-sections.

\subsubsection{2016 PhysioNet Heart Sound Database (PHSDB)}
The 2016 PhysioNet/CinC challenge focused on the task of classifying a heart sound/PCG into the \emph{normal} or \emph{abnormal} class. The PHSDB dataset is consists of eight different subsets {a-g,i} of PCG recording from seven different research groups. Each research group used different stethoscopes to record the PCGs (see Table \ref{data_physionet}). An ensemble of six subsets \{a-f\} has been made public as training data. It consists of 3,153 PCG recordings from 764 patients, a total of 84,425 cardiac cycles. The duration of the recordings varies from 5 seconds to 120 seconds. A test set comprised of subsets \{b-e,g,i\} is kept private for scoring purposes only and not publicly accessible. There are 2488 recordings of Normal heart sound among 3153 recordings, which is almost 80\% of the entire dataset, indicating an imbalance between the \emph{normal} and \emph{abnormal} classes. Some recordings are performed in a clinical setting, while others are in casual environments with various types of noise. We employ the public train dataset in our experiments following the train-test splits used in \cite{prio_vai}.

\subsubsection{DCASE acoustic scene classification dataset 2016}
The DCASE 2016 acoustic scene classification challenge dataset consists of audio samples from 15 (fifteen) different indoor and outdoor locations or environments. These are Beach, Bus, Cafe/Restaurant, Car, City Center, Forest Path, Grocery Store, Home, Library, Metro Station, Office, Park, Residential Area, Train, and Tram. There are 1170 and 390 audio segments in the training and validation dataset, respectively. Each class has 234 samples for training and 78 samples for validation. The 2-channel audio segments are 30 seconds in duration and are recorded in a 24-bit PCM format at a 44.1kHz sampling rate. For our experiments, the DCASE dataset is designed as a four-fold cross-validation task with about 75\% data used for training and the remaining 25\% for validation.


\subsection{Architecture design}\label{architecture_design}
This section describes how the proposed model architecture and hyperparameters are customized for the two different tasks under consideration. In the case of heart sounds, it is well-known that most of the relevant information in the signal lies below 400Hz \cite{heartSoundFreqRange}. Thus, the frequency range for this task for the spectrogram feature extraction layer is set to 0 - 400Hz. We have experimented with different filterbank sizes and found the optimum filter number to be 16. We will call the proposed learnable filterbank layer with 16 filters SpectNet-16. For classification, we used a CNN architecture with three residual blocks, which we refer to as ResNet for convenience, shown in Fig. \ref{fig:model}. This architecture provides the best result trained with pre-extracted mel-scaled spectrogram features.

In the case of the DCASE acoustic scene classification data, it contains many types of sounds ranging from 0 - 44.1kHz. Assuming most of the relevant information lies within 11kHz, we use 46 filters ranging processing inputs in the frequency range of 0-11kHz. A shallower and simpler model is used since larger models cause overfitting for this task. A simple two-layer CNN model is used after the proposed front end layer for this task. This model architecture is inspired by \cite{dcase_super_vector} that showed promising results. This architecture is referred to as the SpectNet-46 + CNN.


\subsection{Training regime}
First, the PHSDB and DCASE datasets are used separately to train the model using fixed parameters in the front-end filterbank layer corresponding to traditional mel-scale spectrograms. In this stage, the front-end of the network is frozen while the 2D CNN backbone model parameters are updated. The entire network is trained in the next stage when the front-end filterbank layer parameters are optimized for task-specific spectrogram feature extraction. In the training process, the categorical cross-entropy loss function is minimized using the Adam optimizer.

For training with the PHSDB, the public training and validation set provided is used. The PHSDB data is recorded using seven different stethoscope devices (data-subset e uses two devices). We refer to the data from these six domains as six different subsets for PHSDB. Since there is an imbalance in the dataset in both classes and subsets, a double balance training (DBT) method \cite{prio_vai} is used to train the model. This method ensures that the training mini-batches are balanced both in terms of classes and domains. The model is trained with a batch size of 996 and a learning rate of 0.001. In contrast to PHSDB, the DCASE 2016 dataset is a balanced dataset that contains equal samples from each class in both train and validation sets. A batch size of 64 and a learning rate of 0.005 is used in this case.


\begin{table*}[hbt!]
    \centering
    \caption{Performance Comparison on DCASE Acoustic Scene Classification Task}
        \begin{tabular}{ccccccc}
\Xhline{2\arrayrulewidth}
Methods            & \textbf{Accuracy} & Sensitivity & Specificity & Precision & F1    & MACC  \\ \Xhline{1\arrayrulewidth}
\multicolumn{7}{c}{\cellcolor[HTML]{EFEFEF}Static SpectNet}                                    \\
SpectNet-16 + CNN  & 66.92             & 65.44       & 97.53       & 65.88     & 65.66 & 81.47 \\
SpectNet-32 + CNN  & 73.59             & 73.59       & 98.11       & 74.37     & 73.98 & 85.85 \\
SpectNet-46 + CNN  & 74.44             & 74.50       & 98.18       & 74.95     & 74.10 & 86.34 \\
SpectNet-64 + CNN  & 73.93             & 73.56       & 98.11       & 74.05     & 73.81 & 85.84 \\
SpectNet-128 + CNN & 75.55             & 75.56       & 98.25       & 76.10     & 75.83 & 86.91 \\
SpectNet-149 + CNN & 74.81             & 74.22       & 98.16       & 74.94     & 74.58 & 86.19 \\
\multicolumn{7}{c}{\cellcolor[HTML]{EFEFEF}Proposed System (Learnable SpectNet)}               \\
SpectNet-46 + CNN  & \textbf{76.55}    & 76.47       & 98.32       & 77.15     & 76.07 & 87.39 \\ \Xhline{2\arrayrulewidth}
\end{tabular}
\label{DcaseTable}
\end{table*}
 
\subsection{Evaluation setup and performance metrics}
We followed the performance metrics instructed by the respective challenge organizers in the experimental evaluation of the Physionet 2016 and DCASE 2016 challenges. The train and validation set provided by the organizers were used to evaluate our system. To measure performance we calculated \emph{sensitivity}, \emph{specificity}, \emph{precision}, \emph{F1 score}, \emph{accuracy} and \emph{Macc} (Modified accuracy - average of sensitivity and specificity) for both datasets. Since Macc was the main metric used comparing different systems in the PHSDB challenge, we focused on Macc for model comparison for the heart sound classification task. In contrast, the accuracy metric is used for the DCASE acoustic scene classification challenge. Therefore, this metric was monitored for comparison in this case.

\subsection{Baseline systems}
Many different approaches were considered for heart sound classification using the PHSDB dataset. However, in most cases, the researchers used custom train-test splits which make it difficult to perform comparative studies \cite{noman2019short, heartsound_mfcc_cnnlstm}. In the original Physionet challenge, the top-performing system was developed by Potes \emph{et al.} \cite{potes}. However, since the reported results were only on the hidden test set, we re-implement this system for evaluation on the public train validation set. This same train-validation set was used in \cite{prio_vai} where the domain-variability problem in heart sound classification was addressed. We replace the tConv filterbank used in \cite{prio_vai} with our proposed SpectNet and use the same model architecture with both 1D and 2D representation from the SpectNet. We denote these systems as Gammatone 1D-CNN and Gammatone 2D-CNN, respectively.

In the case of the DCASE acoustic scene classification task, we use the model architecture used in \cite{dcase_super_vector} and focus on the performance gain achieved by the proposed learnable spectrogram layer. In his work, different features along with the spectrogram image features (SIF) were fused in parallel. Since the proposed system in this work is mainly focused on the spectrogram image extraction, we only compare the results of the SIF system utilized in \cite{dcase_super_vector}. First, we implement the baseline system with the learnable front-end layer keeping it fixed to extract conventional mel-scale spectrogram features. Next, we allow the entire network to learn all the parameters, including the front-end spectrogram layer. The performance comparison is shown in Table \ref{DcaseTable}.

\subsection{Comparison between 1D and 2D feature representations}
In this work, we propose a network with a 2D spectrogram feature front-end layer. However, individual spectrogram frames can also be processed as a sequence of 1D feature vectors. The general hypothesis behind using 2D spectrogram image features is that it would provide improved performance since, in this approach, the network architecture can learn the correlation between different time-frequency regions of the input feature. In this section, we devise an experiment to test this hypothesis on the heart sound classification task.

The baseline system used for PHSDB is \cite{potes}, which comprises four branched 1D-CNNs with 4 predefined filters in the frond-end. We also designed a similar network with four 1D-CNN branches with 4 filters initialized using the mel-frequency scaled center frequencies. We observe similar performance in these two systems. Next, we design an experiment where the outputs of the 4 filters are stacked vertically to construct a 2D feature matrix and then classified using a 2D-CNN model, which has the same depth as the 1D-CNN model used in \cite{prio_vai}. The results in the first two columns of Table \ref{heartSoundTable} show that the system using 2D feature representation provides improved performance in classifying heart sounds compared to the 1D feature representation. Thus, it follows that a learnable spectrogram layer within the network should similarly provide superior performance compared to learnable 1D filterbanks. The final CNN architecture used to classify our 2D feature representation is shown in Fig. \ref{fig:model}.

\subsection{Effect of filterbank size}
In this section, we experiment with the number of filters in our front-end layer. To observe this effect, we first use the PHSDB to run experiments using $4$, $8$, and $16$ filters in the front-end filterbank layer, respectively. The results provided in Table \ref{heartSoundTable} show that the increase of filterbank size did not affect the performance. We believe this is due to the limited bandwidth of PCG signals which is in the range of $0-400$Hz \cite{heartSoundFreqRange}. Thus, only four filters were sufficient to extract the required information from the PCG signal.

Naturally, on the DCASE acoustic scene classification task, the number of filters must be higher for feature extraction since the frequency range is $0-22$kHz, which is far greater than the bandwidth of heart sounds. We use the system in \cite{dcase_super_vector} as our baseline model, which implemented a small model to classify the samples using pre-extracted MFCC features using 149 filterbanks. We use the same model with our proposed layer at the front-end to extract the spectrogram image features in an end-to-end system. We experimented with a various number of filters at the front-end, including the sizes 16, 32, 46, 64, 128, and 149. The results are shown in Table \ref{DcaseTable} where we observe a similar trend, except that the performance plateau is reached when 46 filters are used. We note that increasing the number of filters increases the model size significantly. A decrease in the filter bank size from 46 also resulted in a significant drop in performance, as shown in Table \ref{DcaseTable}. Therefore, we finally selected 46 filters in the front-end for our learnable spectrogram experiments.
 
\subsection{Effect of the learnable filterbank layer}
Finally, we conduct experimental evaluations to determine how the proposed learnable filterbank layer affects the 2D spectrogram feature representation and the final classification performance. For this experiment, we select the best-performing model with the fixed front-end layer for both of our tasks as our starting point. We train the front-end layer with the pre-trained 2D-CNN model with 3 residual blocks for PHSDB. As shown in Table \ref{heartSoundTable}, an improvement of 1.2\% is observed on MACC, the main evaluation metric for PHSDB. The proposed method also improves the other performance metrics, including specificity, precision, and F1 score. To verify that the improvement obtained is statistically significant, we perform the McNemer's test \cite{mcnemer_test} and found that the difference between the baseline system and the proposed system is significant ($p$ \textless  0.05).
 
In the case of the DCASE acoustic scene classification data, the main evaluation metric was \emph{accuracy}. In this case, we use the CNN model used for a fixed-filterbank size of 46 in the previous section as the starting point. In this case, learning the filterbank layer results in an improvement in \emph{accuracy} by $2.11\%$ as evident in Table \ref{DcaseTable}. The proposed method also improves the other evaluation metrics by a significant margin, even surpassing models that use a larger fixed-filter bank. This result shows the effectiveness of the proposed learnable architecture as the model effectively selects the regions of the time-frequency representation that is best suitable for the classification task. We justify this claim by observing that a learnable spectrogram feature of size 46 performs superiorly compared to a fixed spectrogram feature of size 128. This means a set of intelligently selected center frequencies for spectrogram generation is better than a set of pre-defined fixed filters, even if a higher number of filters is used.

\section{Conclusions} \label{conclussion}
This study has developed SpectNet - a deep learning architecture for audio classification that uses a learnable spectrogram feature representation as a front-end layer. 
The proposed method was motivated by the idea that a spectrogram representation with learnable parameters can adapt to the idiosyncratic time-frequency characteristics of the input audio data, which can vary depending on the source domain. Fixed spectrograms and MFCC-based features have already been used widely for audio classification using various 2D CNN architectures and are known to be effective for several tasks. Thus, to demonstrate the proposed method's effectiveness, we have experimented using two different audio classification tasks, namely, heart sound abnormality detection and acoustic scene classification. Customized front-end layers were designed for both tasks based on the optimal size of mel-scaled gammatone filterbanks. Experimental evaluation on the Physionet heart sound classification dataset and the DCASE acoustic scene classification dataset have demonstrated that the proposed learnable front-end layer effectively improves the overall performance of the system for both of these tasks compared to a competitive baseline system. Notably, an improvement of $1.2\%$ in terms of MACC and $2.11\%$ in terms of accuracy has been achieved for the heart sound classification task and acoustic scene classification task, respectively, when compared against competitive architectures using fixed-filterbank spectrogram layers. The favorable experimental results obtained on two different domains thus validate the effectiveness of the proposed method.


\ifCLASSOPTIONcaptionsoff
  \newpage
\fi

\bibliographystyle{IEEEtran}
\bibliography{references.bib}
\end{document}